\documentclass[letterpaper, 10 pt, conference]{ieeeconf}  
\IEEEoverridecommandlockouts                              

\usepackage{graphicx}
\usepackage{cite}
\usepackage{times}
\usepackage{amsmath}
\usepackage{changepage}
\usepackage{amssymb}
\usepackage{amsfonts}
\usepackage{fancyhdr}
\usepackage{comment}
\usepackage{wrapfig}
\usepackage{multirow}\usepackage{multirow}
\usepackage{multicol}\usepackage{multicol}
\usepackage{subfig} 
\usepackage{caption}
\usepackage{morefloats}
\usepackage{siunitx}
\title{\LARGE \bf
	Comparison of control strategies for the temperature control of a refrigeration system based on vapor compression}
\author{Jairo Viola$^{1}$, Alberto Radici$^{2}$, and YangQuan Chen$^{3}$ 
	\thanks{$^{1}$J. Viola is with the Dept. of Mechanical Engineering, School of Engineering, University of California, Merced. 5200 N. Lake Road, Merced, CA 95343, USA. 
		{\tt\small jviola@ucmerced.edu}}%
	\thanks{$^{2}$A. Radici is with the Dept. of Mechanical Engineering, School of Engineering, University of California, Merced. 5200 N. Lake Road, Merced, CA 95343, USA. 
		{\tt\small aradici@ucmerced.edu}}%
	\thanks{$^{3}$Y. Q. Chen is with the Dept. of Mechanical Engineering, School of Engineering, University of California, Merced. 5200 N. Lake Road, Merced, CA 95343, USA. 
		{\tt\small ychen53@ucmerced.edu}}%
}
\begin{document}
	\maketitle
	\thispagestyle{empty}
	\pagestyle{empty}
\begin{abstract}  
This paper presents the design of multivariable temperature control for a refrigeration system based on vapor compression employing the internal model control technique. The refrigeration system is based on the PID18 benchmark, which is a $2\times 2$ MIMO system. The controlled output variables of the refrigeration system are the cooling power managed through the outlet temperature of the evaporator and the superheating degree at the condenser. The input variables of the system are the valve opening and the compressor speed. System identification is performed by applying stepped signals to the input variables, resulting in four transfer functions estimated with a Box-Jenkins model. From the MIMO system transfer functions, the relative gain array is calculated to determinate the best variables to be paired. After that, according to the variables to be paired, the corresponding transfer functions are reduced to order two to design a PID controller for each output variable employing the internal model control technique. Then, the controllers are contrasted employing a set of quantitative performance indexes with the control results achieved in the PID18 workshop. Obtained results show that the proposed Internal Model controllers have better performance than most of the proposed controllers at the PID18.
\end{abstract}
\section{Introduction}
Refrigeration systems are high energy consumption processes, which are employed in many industries as food, medical, aeronautics, comfort among others. According to \cite{c1}, in the United States, the Heat, Ventilating, and Air Conditioning systems consume around the 45\% percent of the energy in commercial and residential buildings. Likewise, in countries like Singapore, with subtropical weather more than the 52\% of the energy produced is consumed by these systems \cite{c2}. Thus, efficient control of the energy consumption of refrigeration systems is required to maintaining the human comfort, the industry requirements, and reduce the environmental and economic impact. Usually, PI and PID controllers are employed for the control of refrigeration systems due to its simplicity for tuning and implementation. For example, \cite{c3} presents the application of a classical PID controller for the control of a critical refrigeration cycle for CO2 as working fluid. Besides, \cite{c4} shows the application of a multivariable decoupled control technique with PI and PID controllers to control a  system. In \cite{c5}, the performance of a split air conditioning system is improved experimentally employing PID controllers. 
On the other hand, there are other control techniques different than PID controllers that can be applied to the control of refrigeration systems. In \cite{c6}, a neural network is combined with a gain-scheduled PI-based controller for the superheating control in a refrigeration system. Also, an $H_{\infty}$ control of a refrigeration system is proposed by \cite{c7}, which considers many operating points as a multivariable controllability analysis of the system. Also, \cite{c2} presents the application of MPC based controller for the temperature control of a vapor compression refrigeration system to minimize the coefficient of performance (COP).\par
This paper presents the design of multivariable temperature control of a refrigeration system based on vapor compression employing the internal model control technique with PID controllers. The refrigeration system is based on the PID18 benchmark model \cite{c1}, which is modeled as a $2\times 2$ multiple input multiple output system (MIMO). The system outputs are the evaporator output temperature and the overheating temperature in the condenser. The model inputs are the opening percentage of the expansion valve and the compressor speed.
The identification process is performed by applying stepped profile signals to the system inputs between the allowed variation ranges to obtain four transfer functions that represent the dynamic behavior of the refrigeration system. Besides, the Box-Jenkins model is employed to define the system transfer functions. 
From the identified transfer functions, the relative gain array (RGA) is calculated, to determinate the best pairing variables of the system to be controlled. The RGA matrix shows that the expansion valve opening controls the evaporator outlet temperature, and the compressor speed controls the overheating temperature in the condenser. After that, the paired transfer functions are reduced to a second order model to design a PID controller for each output variable employing the internal model control technique (IMC) proposed by \cite{c8}. Then, the obtained PID-IMC controllers performance is contrasted with a decentralized PID controller proposed by the  PID18 workshop creators, employing a set of quantitative performance indexes. After that, the PID-IMC controllers are readjusted to obtain an improved control response of the system with respect to the decentralized PID controller. Finally, the obtained results for the PID-IMC controllers are contrasted with all the proposed control strategies for the refrigeration system presented during the PID18 conference.\par
The main contribution of this paper is the use of an internal model control strategy for the temperature control of the refrigeration system proposed on the PID18 benchmark, and a performance comparison with all the control strategies proposed during the PID18 benchmark challenge. The obtained results show that the proposed PID-IMC controller has a good performance similar to the one obtained employing advanced control strategies, with several benefits as an easy practical implementation.
This paper is structured as follows. First, the refrigeration system based on vapor compression is described. Second, the system identification and the RGA matrix are calculated. Third, the PID-IMC controllers are tuned and contrasted with the decentralized PID controllers proposed by PID18 benchmark creators. Fourth, the PID-IMC controller improvement and its comparison with all the control strategies proposed during PID18 challenge is developed. Finally, the conclusions are presented.
\section{Refrigeration system based on vapor compression}
The model of the refrigeration system based on vapor compression proposed at the benchmark PID18 workshop is presented in Fig.1 \cite{c1,c9}. As can be observed, this is a closed loop system composed by a compressor, an evaporator, and the system actuators, which are the expansion valve with an opening percentage $A_{v}$ and the compressor with speed $N$. 
Also, the parameters that describe the refrigeration system dynamical behavior are the evaporator output temperature on the secondary flux $T_{e,sec,out}$, the super heating degree $T_{sh}$, the mass flow in the secondary fluid of the evaporator $\dot{M}_{e,sec}$, the mass flow in the secondary fluid of the condenser $\dot{M}_{c,sec}$, the inlet temperature of the secondary flow in the condenser $T_{c,sec,in}$, and the inlet temperature of the secondary flow in the evaporator $T_{e,sec,in}$.
\begin{figure}
		\centering
			\includegraphics[width=7cm]{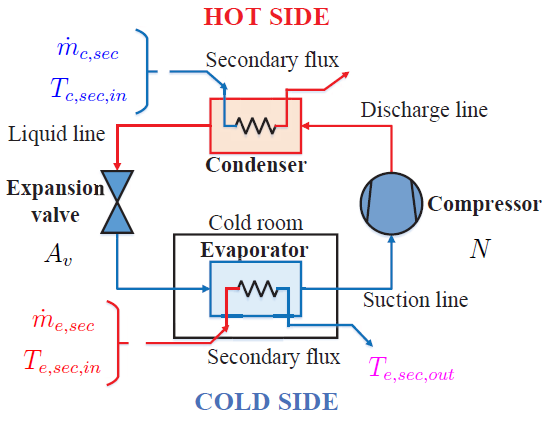}
			\caption{Refrigeration system based on vapor compression }
			\label{fig1} 
\end{figure}
According to \cite{c1}, in the closed loop refrigeration system, the circulating refrigerant passes through the compressor as saturated vapor, compressing it into super heated vapor with higher pressure and temperature. Then, the super heated vapor is sent to the condenser, where the refrigerant flows through coil pipes, cooling the vapor and condensing it into a liquid phase. After that, the condensed vapor pressure is reduced abruptly with the expansion valve. Next, the expanded vapor enters in the evaporator, where the fluid evaporates, and the environment air heat is transferred to the fluid.  
For the proposed refrigeration system, the control goals are satisfying the cooling demand $\dot{Q_{e}}$ represented by the evaporator output temperature on the secondary flux $T_{e,sec,out}$, as well as the super heating degree $T_{sh}$ to improve the cooling power efficiency, measured through the COP index employed on refrigeration systems. This index is defined as the ratio of cooling demand and the mechanical power of the compressor $\dot{W_{comp}}$.
Notice the $\dot{m}_{e,sec}$, $\dot{m}_{c,sec}$, $T_{e,sec}$, $T_{c,sec}$, $T_{e,sec,in}$ and $T_{c,sec,in}$ act only as measurable disturbances. Thus, the only controlled variables are $A_{v}$ and $N$. It means the refrigeration system can be modeled as a $2\times 2$ MIMO system. Table I and Table II present the control variables and the disturbances applied to the refrigeration system. As can be observed, Table I and Table II define the initial condition for the manipulated variables and disturbances that define the dynamical behavior of the system. In addition, Table II considers as disturbances the input pressure of the secondary flow at the condenser and the evaporator  $T_{c,sec,int}$ and $T_{e,sec,int}$ respectively as the compressor surrounding pressure $T_{surr}$.
\begin{table}
		\caption{Refrigeration system inputs and outputs}
		\centering
			\begin{tabular}{|c|c|c|c|}		
				\hline
				& \multirow{2}{*}{\centering Parameter} & \multirow{2}{*}{\centering Range} & \multirow{1}{*}{Initial}\\
				& & &condition\\
				\hline
				\multirow{1}{*}{System}&$N$& 30-50 Hz&40Hz \\
				inputs & $A_{v}$& 10\% to 90\%&50\%\\
				\hline
				\multirow{1}{*}{System}&$T_{sec,out}$&$-22.1^{\circ}C$- $-22.6^{\circ}C$&$-22.1^{\circ}C$\\
				 Outputs & $T_{sh}$&$7.2^{\circ}C$ to $22.2^{\circ}C$&$14.65^{\circ}C$\\
				\hline
			\end{tabular}
	\end{table}
\begin{table}
		\caption{Refrigeration system disturbances}
		\centering
			\begin{tabular}{|c|c|}	
				\hline
				Parameter & Initial condition\\
				\hline
				$T_{c,sec,in} $ & $30^{\circ}C$\\
				\hline
				$\dot{m}_{c,sec}$ & $150$ $gs^{-1}$\\
				\hline
				$P_{c,sec,in} $ & $1$ $bar$\\
				\hline
				$T_{e,sec,in} $ & $20^{\circ}C$\\
				\hline
				$\dot{m}_{e,sec} $ & $1$ $bar$\\
				\hline
				$T_{surr}$ & $25^{\circ}C$\\
				\hline
			\end{tabular}
\end{table}
\section{Refrigeration system identification}
This section presents the system modeling as a $2\times 2$ MIMO system, the identification process employing transfer functions estimated with the Box-Jenkins method, and the evaluation of the relative gain array to find the best pairing variables.
\subsection{Multivariable modeling}
The refrigeration system can be represented as the multivariable transfer function model shown in Fig.2. As can be observed the input variables are the expansion valve opening percentage $A_{v}$ and the compressor speed $N$, and the output variables are the $T_{e,sec,out}$, and the $T_{sh}$. The linear transfer functions $G_{11}$, $G_{12}$, $G_{21}$, and $G_{22}$ represent the relationships between the system input and output variables, where the first number is the output and the second number the input. For example, the transfer function $G_{12}$ represent the behavior of the $T_{sh}$ against $A_{v}$.
	\begin{figure}
	\centering
			\includegraphics[width=4cm,height=3cm]{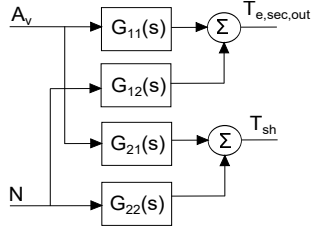}
			\caption{MIMO transfer function model for the refrigeration system} 
	\end{figure}
\subsection{System identification}
For the multivariable model proposed for the refrigeration system presented in Fig.2, the identification process must find the transfer functions $G_{11}$, $G_{12}$, $G_{21}$, and $G_{22}$ that represent the dynamic behavior of the system, and allows finding the best pairing variables for the controller design. For this reason, the identification process first set a stepped signal on the valve opening $A_{v}$, keeping the compressor speed constant to 40 Hz, or its initial condition to obtain the system response of $T_{e,sec,out}$, and $T_{sh}$ required to identify $G_{11}$ and $G_{21}$. Then, the stepped signal is set for the compressor speed input with the valve opening on 50\% to identify $G_{12}$ and $G_{22}$. Figure 3 shows the stepped signals applied to $A_{v}$ and $N$, and the response obtained for $T_{e,sec,out}$, and $T_{sh}$.
\begin{figure}
\centering
\subfloat[]{\includegraphics[width=0.25\textwidth,height=0.12\textheight]{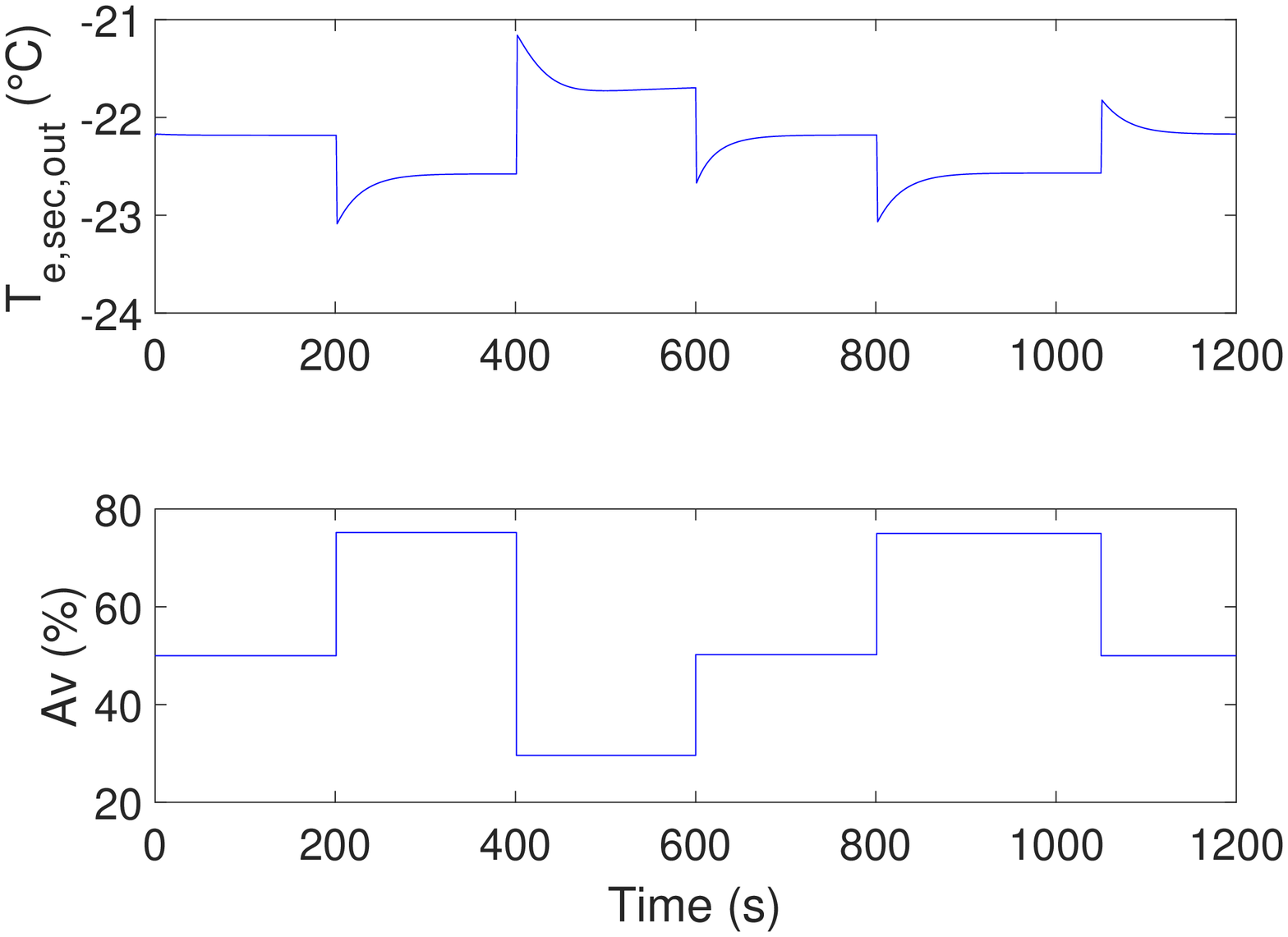}} 
\subfloat[]{\includegraphics[width=0.25\textwidth,height=0.12\textheight]{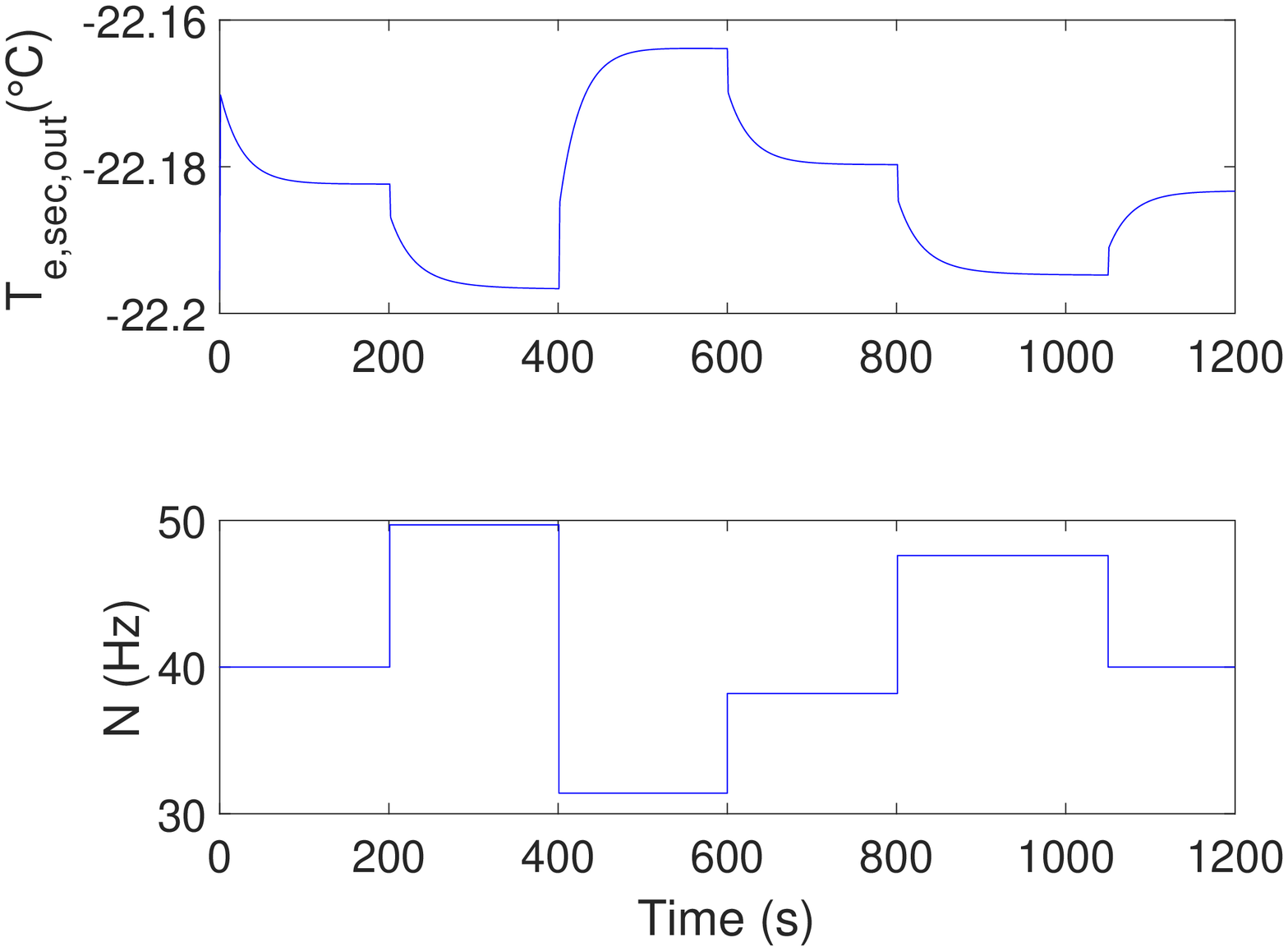}}\\
\noindent 
\subfloat[]{\includegraphics[width=0.25\textwidth,height=0.12\textheight]{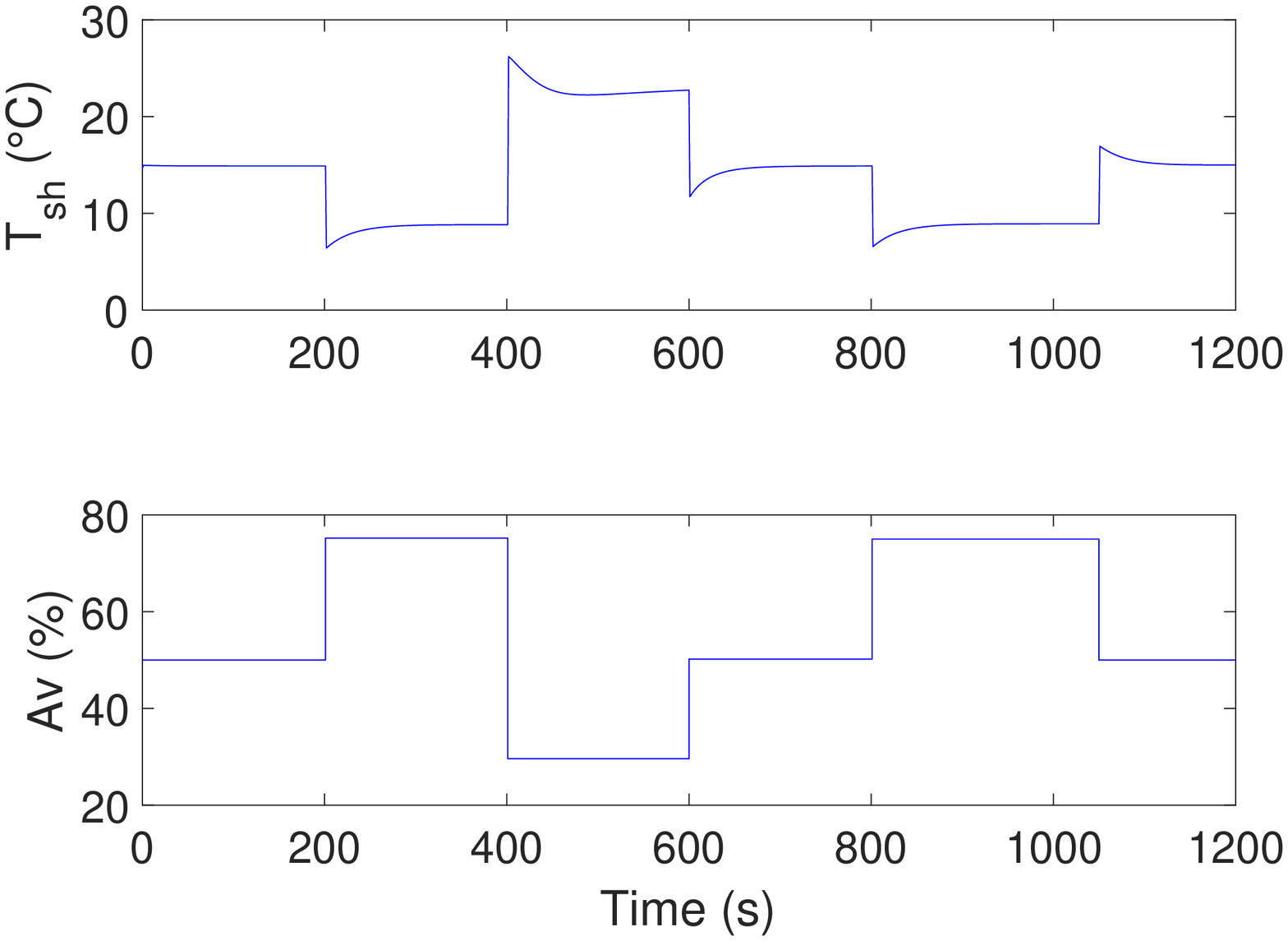}} 
\subfloat[]{\includegraphics[width=0.25\textwidth,height=0.12\textheight]{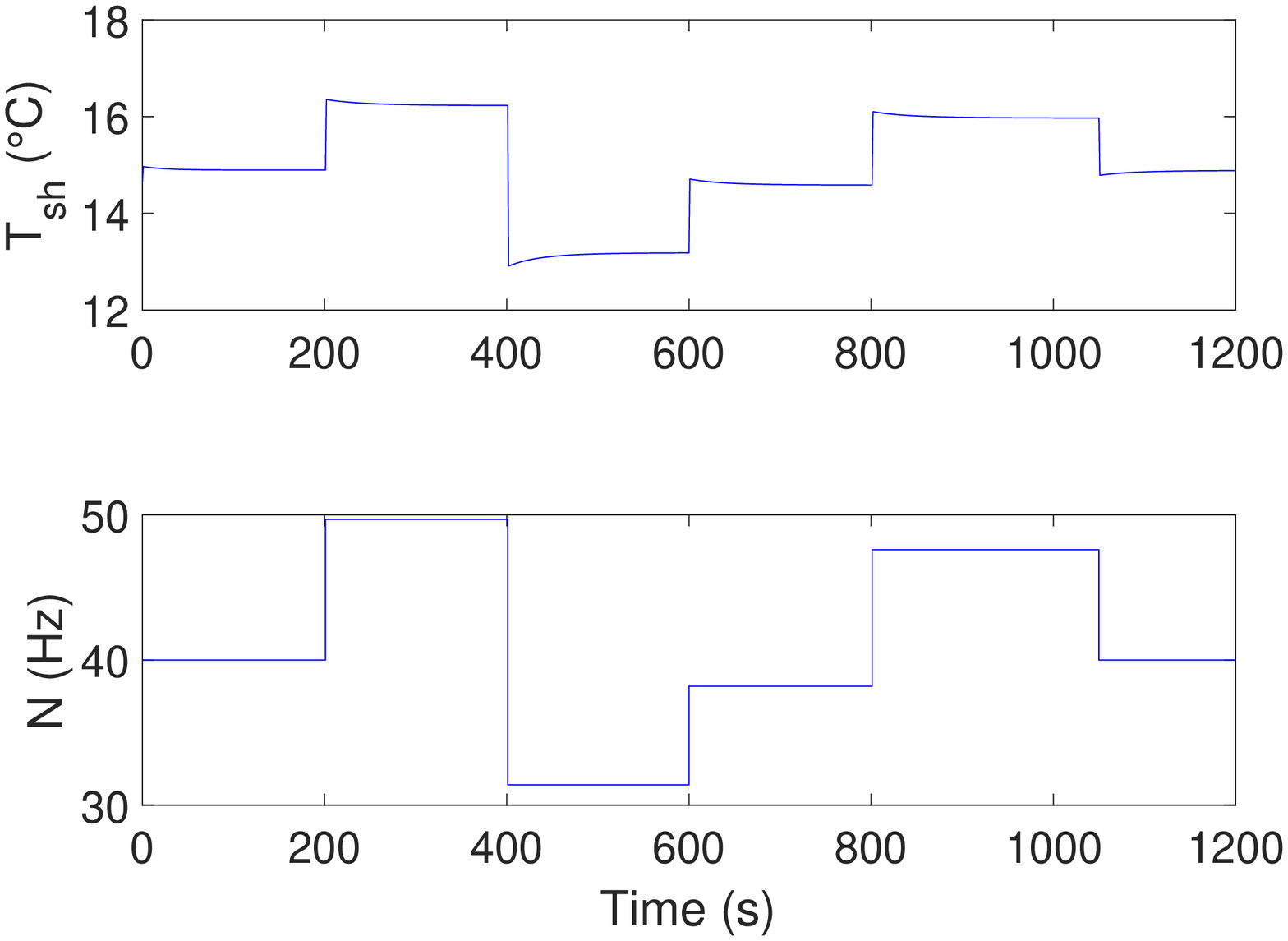}}
\caption[caption]{Stepped identification signals for a) $G_{11}$ b) $G_{12}$ c) $G_{21}$ d) $G_{22}$}
\label{fig2}
\label{fig3}
\end{figure}
The next step in the identification process is to find a linear model of the refrigeration system with the data obtained from the PID18 benchmark model. According to \cite{c1}, the Box-Jenkins method (BJ) give the best representation of the dynamical behavior of the refrigeration system. Figure 4 shows the block diagram of a BJ model \cite{c10}. As can be observed, the model consists of two ratios. The ratio $\frac{B(z)}{A(z)}$ represents numerator and denominator of the discrete transfer function of the system. On the other hand, the ratio $\frac{C(z)}{D(z)}$ represents the numerator and denominator of the transfer function of the disturbances introduced to the system. Employing the Matlab Identification Toolbox for the data presented in Fig.3, the discrete BJ transfer functions $G_{11}$, $G_{12}$, $G_{21}$, and $G_{22}$ of the refrigeration system are given by (1). For all the identified transfer functions, the fitness value is about 87\%. In the case of $G_{11}$ and $G_{21}$ the BJ model employs a second order polynomial for the numerator and denominator, and for $G_{12}$ and $G_{22}$ a fourth order polynomial for the system representation.
\begin{figure}
	\centering
		\includegraphics[width=4cm,height=2.5cm]{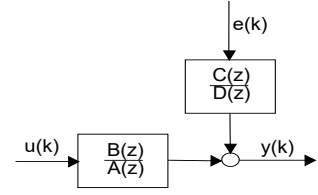}
		\caption{Box-Jenkins model}
\label{fig4}
\end{figure}
\begin{eqnarray}
\begin{scriptsize}
\left.\begin{aligned}
G_{11}=&\frac{-0.03408z^{-1}+0.03357z^{-2}}{1-0.9699z^{-1}+0.001037z^{-2}}\\
G_{12}=&\frac{-0.00006045z^{-1}+0.00075z^{-2}+0.0002z^{-3}-0.0003z^{-4}}{1-1.298z^{-1}-0.344z^{-2}+0.64z^{-3}-0.0024z^{-4}}\\
G_{21}=&\frac{-0.3765z^{-1}+0.3706z^{-2}}{1-0.9775z^{-1}+0.000528z^{-2}}\\
G_{22}=&\frac{0.1746z^{-1}-0.1639z^{-2}-0.1744z^{-3}+0.1637z^{-4}}{1-0.9375z^{-1}-0.9976z^{-2}+0.9367z^{-3}-0.001551z^{-4}}
\end{aligned}\right.
\end{scriptsize}
\end{eqnarray}
\subsection{RGA matrix calculation}
The RGA matrix is a useful tool for multivariable systems, which allows finding the relationship between the system variables, and its best pairing for the controller design. The RGA matrix is calculated based on the steady state array of $G_{11}$, $G_{12}$, $G_{21}$ and $G_{22}$ when $z\rightarrow1$ given by (2).
\begin{equation}
\left.\begin{aligned}
\raisebox{\depth}{$\displaystyle 
A=\begin{bmatrix}
G_{11}|_{z\rightarrow 1} & G_{12}|_{z\rightarrow 1}\\
G_{21}|_{z\rightarrow 1} & G_{22}|_{z\rightarrow 1}\\
\end{bmatrix}$}.
\end{aligned}\right.
\end{equation}
The RGA matrix if the system is calculated using (3), and the steady state array $A$. For the identified refrigeration system, the RGA matrix is given by (4)
\begin{equation}
RGA=A\times(A^{-1})^T.
\end{equation}
\begin{equation}
\left.\begin{aligned}
\raisebox{\depth}{$\displaystyle RGA=\begin{bmatrix}
1.0004 & -0.0004\\
-0.0004 & 1.0004\\
\end{bmatrix}$}.
\end{aligned}\right.
\end{equation}
As can be observed, the RGA matrix is almost 1 in the main diagonal. So that, it can be said that $G_{11}$ relates the outlet evaporator temperature $T_{e,sec,out}$ with the expansion valve opening percentage $A_{v}$, and $G_{22}$ relates the superheating temperature $T_{sh}$ with the compressor speed $N$. Therefore, $G_{11}$ and $G_{22}$ are the required transfer functions for the multivariable control system design.
\section{Refrigeration system control}
This section presents the design of the PID controllers for the control of $T_{e,sec,out}$ and $T_{sh}$ employing the internal model control technique based on the identified MIMO model of the system found in Section III.
\subsection{PID controller}
According to \cite{c11}, a PID controller is represented in the Laplace domain by the transfer function (5), where $U(s)$ is the control action, $E(s)$ the feedback system error, $k$ the proportional term, $\tau_{i}$ the integral time term, $\tau_{d}$ the derivative time term, and $N$ is the filter coefficient of the derivative action.
\begin{equation}
	\left.\begin{aligned}
	\raisebox{\depth}{$\displaystyle PID(s)=\frac{U(s)}{E(s)}=k\left(1+\frac{1}{\tau_{i}s}+\frac{\tau_{d}s}{\frac{\tau_{d}s}{N}+1}\right)$}.
	\end{aligned}\right.
\end{equation}
\subsection{Internal model control}   
The internal model control technique (IMC) is based on the internal model principle which proposes that a precise control of a system can be reached only if the control system considers the model of the process to be controlled. According to \cite{c8}, the IMC control structure is given by (6):
\begin{equation}
	\left.\begin{aligned}
	IMC(s)=\frac{f(s)G_{p}^{-1}}{1-f(s)\tilde{G_{p}}^{-1}\tilde{G_{p}}}
	\end{aligned}\right.
\end{equation}
where $\tilde{G_{p}}$ is the identified model of the system, $\tilde{G_{p}}^{-1}$ is the invertible term of the system, and $f(s)$ is the controller filter given by (7):
\begin{equation}
	\left.\begin{aligned}
	f(s)=\frac{1}{(\lambda s+1)^{n}}
	\end{aligned}\right.
\end{equation}
where $\lambda$ is a tuning parameter and $n$ is the filter order, which is selected to ensure that (6) remain as proper transfer function. If $G_{p}$ is a second order system with real poles given by (8)
\begin{equation}
	\left.\begin{aligned}
	G_{p}=\frac{k_{p}}{(\tau_{1}s+1)(\tau_{2}+1)}
	\end{aligned}\right.
\end{equation}
where $k_{p}$ is the plant gain, and $\tau_{1}$, $\tau_{2}$ are the time constants of the system. Replacing (7) and (8) in (6), and considering a filter order $n=1$, the resulting controller can be arranged as a PID controller (PID-IMC) whose proportional, integral, and derivative terms can be calculated using (9).
\begin{eqnarray}
	\left.\begin{aligned}
	k=\frac{\tau_{1}+\tau_{2}}{\lambda k_{p}}\\
	\tau_{i}=\tau_{1}+\tau_{2}\\
	\raisebox{\depth}{$\displaystyle \tau_{d}=\frac{\tau_{1}\tau_{2}}{\tau_{1}+\tau_{2}}$}.
	\end{aligned}\right.
\end{eqnarray}
Notice that in (9), $\tau_{i}$ and $\tau_{d}$ are in function of system model parameters only, while the proportional constant $k$ depends not only of system model but also of $\lambda$, which defines the closed loop performance of the system.
\subsection{Model reduction}
According to the design procedure of the PID controllers employing the IMC technique, a second order with real poles model is required to calculate the proportional, integral, and derivate terms of the PID controllers for $G_{11}$ and $G_{22}$. Therefore, both transfer functions should be order-reduced to apply the IMC technique. In (10), the  Laplace domain response of $G_{11}$ and $G_{22}$ is presented. Then, its step response is calculated, and two second order with real poles transfer functions (11) are calculated for $G_{11}$ and $G_{22}$ employing the Matlab System Identification Toolbox with a fit of 98\% denoted by $G_{11red}$ and $G_{22red}$.\\
\begin{eqnarray}
	\begin{scriptsize}
	\left.\begin{aligned}
	G_{11}=&\frac{-0.2367s-0.0035}{s^2+6.872s+0.1269}\\
	G_{22}=&\frac{1.122s^4+0.07231 s^3+11.08 s^2+0.704s+3.712x10^{-6} }{s^5+6.469 s^4+10.3 s^3+63.84 s^2+4.22s-6.288x10^{-8}}
	\end{aligned}\right.
	\end{scriptsize}
	\end{eqnarray}
	\begin{eqnarray}
	\left.\begin{aligned}
	G_{11red}=\frac{-0.016}{(31s+1)(0.00003s+1)}\\
	\raisebox{\depth}{$\displaystyle {G_{22red}}=\frac{0.16}{(3s+1)(0.0000001s+1)}$}.
	\end{aligned}\right.
	\end{eqnarray}
\subsection{PID-IMC controllers design}
The design of the PID-IMC controllers depends of the selection of the value of $\lambda$ as shown in (9), which is the only tuning parameter of the PID-IMC controller, and should be adjusted to get a desired closed loop system response like the obtained in \cite{c1}. Considering the refrigeration system as a multivariable system defined by $G_{11red}$ and $G_{22red}$, two values of $\lambda$ are required, $\lambda_{11}$ for $G_{11red}$, and $\lambda_{22}$ for $G_{22red}$, which are founded experimentally based on the closed loop system response. Equating (9) for $G_{11red}$  with $\lambda_{11}=0.2$, and equating (9) for $G_{22red}$ with $\lambda_{22}=0.2$, the proportional, integral, and derivative terms for both PID-IMC controllers are presented in Table III. Notice the derivative filter coefficient in (5) is set as $N=10$ for each controller. The saturation limits for each controller are the same presented in Table I for the manipulated variables. In addition, the  anti windup back-calculation scheme is combined with the PID-IMC controllers considering the anti windup constant as $\sqrt{\tau_{i}\tau_{d}}$ as proposed by \cite{c11}.
\begin{table}
		\caption{PID-IMC controllers parameters}
		\centering
			\begin{tabular}{|c|c|c|c|}	
				\hline
				Controller & $k$ & $\tau_{i}$ & $\tau_{d}$ \\
				\hline
				PID-IMC $G_{11}$ &\num{-1.93e4}& $31$ & \num{3e-5}\\
				\hline
				PID-IMC $G_{22}$ & $187.5$ & $3$ & \num{1e-7}\\
				\hline
			\end{tabular}
	\end{table}
\subsection{Obtained results}
The PID-IMC $G_{11}$ and PID-IMC $G_{22}$ controllers for the refrigeration system with the output variables $T_{e,sec,out}$ and $T_{sh}$ are tested in the presence of the system disturbances related in Table II. Moreover, the PID-IMC controllers are compared with the multivariable decentralized PID controllers proposed in \cite{c1} for the refrigeration system. Figure 5 shows the time response of the output variables $T_{e,sec,out}$ and $T_{sh}$ of the PID-IMC controllers against the multivariable decentralized PID controllers. As can be observed, the refrigeration system reaches the desired operating conditions with the  PID-IMC controllers for the setpoint defined in benchmark PID18. The performance of the control action for the expansion valve $A_{v}$ and the speed compressor $N$ is shown in Fig.\ref{fig6}. As can be observed, the control action for the PID-IMC controllers has a better transient response than the decentralized PID controllers, specially for the compressor speed with a small peak value at $t=16$ minutes. Regarding to the expansion valve, the response is similar for both controllers. On the other hand, Fig. 7 shows the compressor energy efficiency and the coeficient of performance (COP) for the PID-IMC and the decentralized controllers. As can be observed, the PID-IMC and the decentralized controllers have a similar compressor efficiency and COP except in $t=16$ minutes, where the PID-IMC controllers has a faster response than the decentralized controller.
\begin{figure}
		\centering
			\includegraphics[width=8cm,height=5cm]{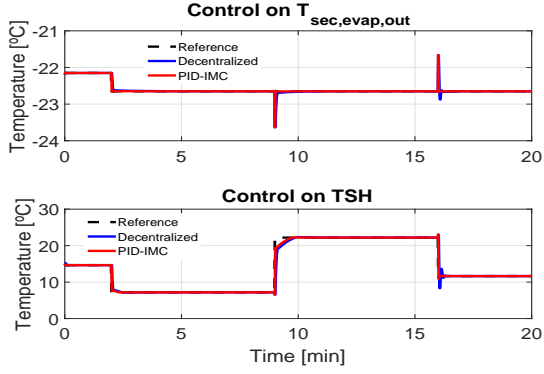}
			\caption{Time response of $T_{e,sec,out}$ and $T_{sh}$ for the decentralized PID and the PID-IMC controllers} 
			\label{fig5}
\end{figure}
\begin{figure}
		\centering
		\includegraphics[width=8cm,height=5cm]{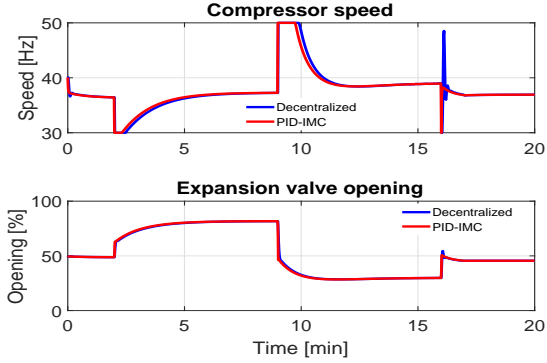}
		\caption{Compressor speed and expansion opening valve for for the decentralized PID and the PID-IMC controllers} 
		\label{fig6}
\end{figure}
\begin{figure}
		\centering
			\includegraphics[width=8cm,height=5cm]{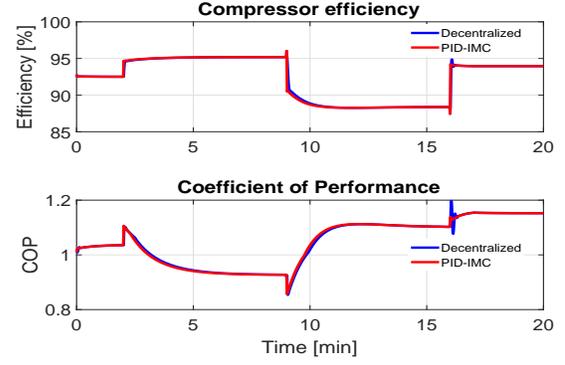}
			\caption{Compressor efficiency and COP for the decentralized PID and the PID-IMC controllers}
			\label{fig7}
\end{figure}
\section{Results analysis}
The set of performance indices (12) is proposed in \cite{c1} to perform a quantitative analysis of the controllers employed in the refrigeration system. These indices measure the ratio of the integral absolute error ($RIAE$), the ratio of integral time absolute error ($RITAE$), and the ratio of integral absolute variation of the control signal ($RIAVU$) between the decentralized PID controllers and the PID-IMC controllers for all the trajectory and during the step transients. In addition, a performance index $J(c_{2},c_{1})$ is considered to measure the global performance of the system. Notice that the weights $w_{i}$ on $J$ are defined by the benchmark creators and are unknown for the benchmark users. Table IV presents the performance indices obtained for the decentralized PID controllers and the PID-IMC controllers. As can be observed, the $RIAE_{1}(C_{2},C_{1})$ and $RIAE_{2}(C_{2},C_{1})$ values are lower for the PID-IMC controllers, indicating better performance during all the trajectory. Likewise, PID IMC controllers has lower $RITAE_{1}(C_{2},C_{1},t_{c2},t_{s1})$, $RITAE_{2}(C_{2},C_{1},t_{c2},t_{s2})$, $RITAE_{2}(C_{2},C_{1},t_{c3},t_{s3})$, and $RITAE_{2}(C_{2},C_{1},t_{c4},t_{s4})$  values than decentralized PID controllers showing that PID IMC controllers have a better response in the presence of the trajectory changes for the $T_{e,sec,out}$ and $T_{sh}$, which are produced by the system disturbances.  Regarding to the $RIAVU_{1}(C_{2},C_{1})$ for $T_{e,sec,out}$, it can be observed that the control action of the PID-IMC controller is higher than decentralized PID controller; however, it improves the PID-IMC controller response as related by the other performance indices. In the case of $RIAVU_{2}(C_{2},C_{1})$ for the $T_{sh}$, there is an improvement for the PID-IMC controller over the decentralized controller indicating lower energy consumption in the compressor. Finally, the overall performance index $J(C_{2},C_{1})$ is lower for the PID-IMC controllers than the decentralized PID controllers. Therefore, it is possible to say that the PID-IMC controllers a have better performance for the temperature control of the refrigeration system during tracking tasks and in the presence of external disturbances.
\begin{eqnarray}
\begin{scriptsize}
\left.\begin{aligned}
IAE = \int_{0}^{time} |e_{i}(t)| \cdot dt\\
ITAE = \int_{t_{c}}^{t_{c}+t_{2}} (t-t_{c})|e_{i}(t)| \cdot dt\\
IAVU_{i}=\int_{0}^{time} |\frac{du_{i}(t)}{dt}| \cdot dt\\
RIAE_{i}(c_{2},c_{1})=\frac{IAE_{i}(C_{2})}{IAE_{i}(c_{1})}\\
RITAE_{i}(C_{2},C_{1},t_{c2},t_{s1})=\frac{ITAE_{i}(C_{2})}{ITAE_{i}(c_{1})}\\
RIAVU_{i}(c_{2},c_{1})=\frac{IAVU_{i}(C_{2})}{IAVU_{i}(c_{1})}\\
J(c_{2},c_{1})=\frac{1}{\sum_{1}^{8}w_{i}}[w_{1}RIAE_{1}(c_{2},c_{1})\\
+w_{2}RIAE_{2}(c_{2},c_{1})\\
+w_{3}RITAE_{1}(C_{2},C_{1},t_{c1},t_{s1})\\
+w_{4}RITAE_{2}(C_{2},C_{1},t_{c2},t_{s2})\\
+w_{5}RITAE_{2}(C_{2},C_{1},t_{c3},t_{s3})\\
+w_{6}RITAE_{2}(C_{2},C_{1},t_{c4},t_{s4})\\
+w_{7}RIAVU_{1}(c_{2},c_{1})\\
+w_{8}RIAVU_{2}(c_{2},c_{1})]
\end{aligned}\right.
\end{scriptsize}
\end{eqnarray}

\begin{table}
	\caption{Performance indices for the decentralized and PID-IMC controllers}
	\centering
		\begin{tabular}{|c|c|c|}	
			\hline
			\multirow{2}{*}{Index} & Decentralized  & PID\\
			&PID  & IMC \\
			\hline
			$RIAE_{1}(C_{2},C_{1})$ & 0.3511 & 0.076 \\
			\hline
			$RIAE_{2}(C_{2},C_{1})$& 0.4458 & 0.2052\\
			\hline
			$RITAE_{1}(C_{2},C_{1},t_{c2},t_{s1})$& 1.6104 & 0.0411\\
			\hline
			$RITAE_{2}(C_{2},C_{1},t_{c2},t_{s2})$& 0.1830 & 0.0163\\
			\hline
			$RITAE_{2}(C_{2},C_{1},t_{c3},t_{s3})$& 0.3196 & 0.1195\\
			\hline
			$RITAE_{2}(C_{2},C_{1},t_{c4},t_{s4})$& 0.1280 & 0.0051\\
			\hline
			$RIAVU_{1}(C_{2},C_{1})$ & 1.1283 & 3.02085 \\
			\hline
			$RIAVU_{2}(C_{2},C_{1})$ & 1.3739 & 1.1098\\
			\hline
			$J(C_{2},C_{1})$ & 0.68209 & 0.2163 \\
			\hline
		\end{tabular}
\end{table}
\subsection{PID-IMC controllers improvement}
Considering that $\lambda_{11}$ and $\lambda_{22}$ values are chosen manually according to the performance  obtained in the simulations of refrigeration system, the search of an optimal value of $\lambda_{11}$ and $\lambda_{22}$ allows obtaining better performance of the refrigeration system employing the PID-IMC controllers. For this reason, the simulation of the refrigeration system is performed for different values of $\lambda_{11}$ and $\lambda_{22}$ in the range of $0.01\le \lambda_{11} \le 0.51$, and $0.01 \le\lambda_{22}\le 0.51$ with steps of 0.05 for a total of 100 simulations of the system. The choose of this range of values for $\lambda_{11}$ and $\lambda_{22}$ is because $\lambda$ values higher than 0.5 results into a higher values of $J$ index, and for $\lambda$ values lower than 0.01 the system simulation exhibit oscillatory behavior. Indeed, the best values of $\lambda_{11}$ and $\lambda_{22}$ are chosen considering which values minimize the $J$ index performance.
Figure 8 shows the resulting $J$ index in function of $\lambda_{11}$ and $\lambda_{22}$ values. As can be observed, the $J$ index surface has a linear shape, and shows that higher values of $\lambda_{11}$ and $\lambda_{22}$ are related with higher $J$ index values. A rapid inspection of the surface shows that the minimum value of $J$ is reached for $\lambda_{11}=0.01$ and $\lambda_{22}=0.11$. Table V shows the performance indices for the new values of $\lambda_{11}$ and $\lambda_{22}$, which are smaller than the values obtained in Table IV.
\begin{figure}
		\centering
		\includegraphics[width=7cm,height=3cm]{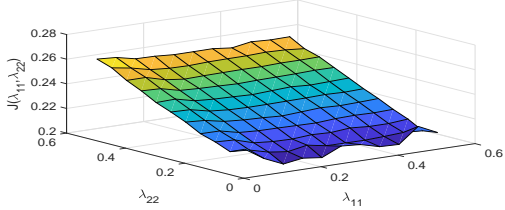}
		\caption {$J$ index surface for different values of $\lambda_{11}$ and $\lambda_{22}$} 
		\label{fig8}
\end{figure}
\begin{table}
	\caption{Performance indices for the decentralized and improved PID-IMC controllers}
	\centering
		\begin{tabular}{|c|c|c|c|}	
			\hline
			\multirow{2}{*}{Index} & Decentralized  & PID  & Improved \\
			&PID  & IMC & PID-IMC\\
			\hline
			$RIAE_{1}(C_{2},C_{1})$ & 0.3511 & 0.076  & 0.071  \\
			\hline
			$RIAE_{2}(C_{2},C_{1})$& 0.4458 & 0.2052  & 0.20 \\
			\hline
			$RITAE_{1}(C_{2},C_{1},t_{c2},t_{s1})$& 1.6104 & 0.04  & 0.0003 \\
			\hline
			$RITAE_{2}(C_{2},C_{1},t_{c2},t_{s2})$& 0.1830 & 0.0163  & 0.0161 \\
			\hline
			$RITAE_{2}(C_{2},C_{1},t_{c3},t_{s3})$& 0.3196 & 0.1195  & 0.1161 \\
			\hline
			$RITAE_{2}(C_{2},C_{1},t_{c4},t_{s4})$& 0.1280 & 0.0051  & 0.003 \\
			\hline
			$RIAVU_{1}(C_{2},C_{1})$ & 1.1283 & 3.02085  & 3.12\\
			\hline
			$RIAVU_{2}(C_{2},C_{1})$ & 1.3739 & 1.11  & 1.10\\
			\hline
			$J(C_{2},C_{1})$ & 0.68209 & 0.216  & 0.2065\\
			\hline
		\end{tabular}
\end{table}
On the other hand, Fig. {\ref{fig9}} to Fig. {\ref{fig12}} shows the surfaces obtained for the other performance indices in function of $\lambda_{11}$ and $\lambda_{22}$. In the case of the RIAE index for the controller PID-IMC $G_{11}$, Fig. 9a shows linear shape surface, where the point $\lambda_{11}=0.01$ and $\lambda_{22}=0.11$ is in the bottom, indicating that the controller have its best performance.  For the $RIAE$ index of the controller PID-IMC $G_{11}$, Fig. 9b shows that even with a less linear surface, the point $\lambda_{11}=0.01$ and $\lambda_{22}=0.11$ has the lowest value for this index.
\begin{figure}
	\begin{center}
		\subfloat[]{\includegraphics[width=0.25\textwidth,height=0.09\textheight]{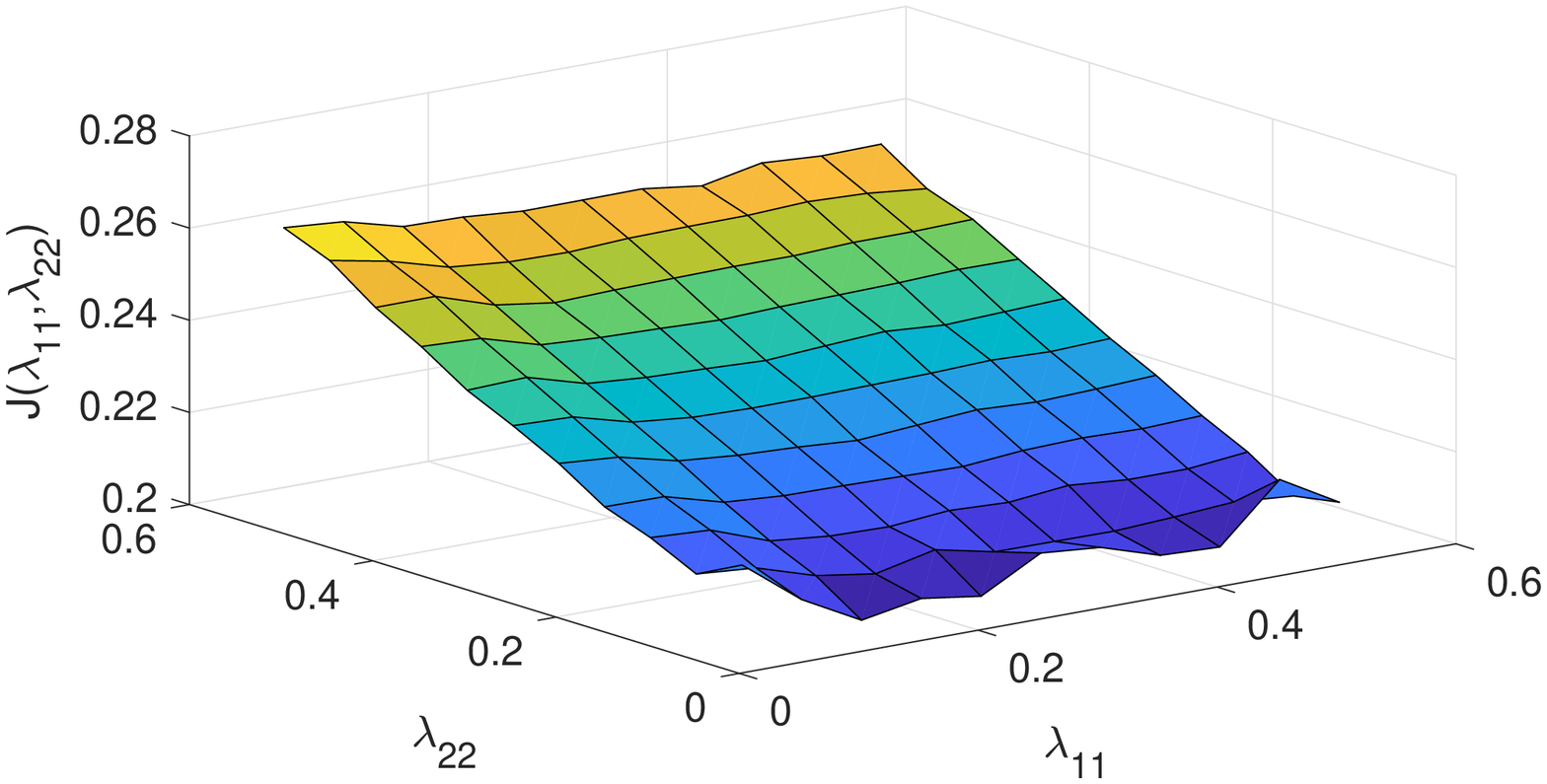}} 
		\subfloat[]{\includegraphics[width=0.25\textwidth,height=0.09\textheight]{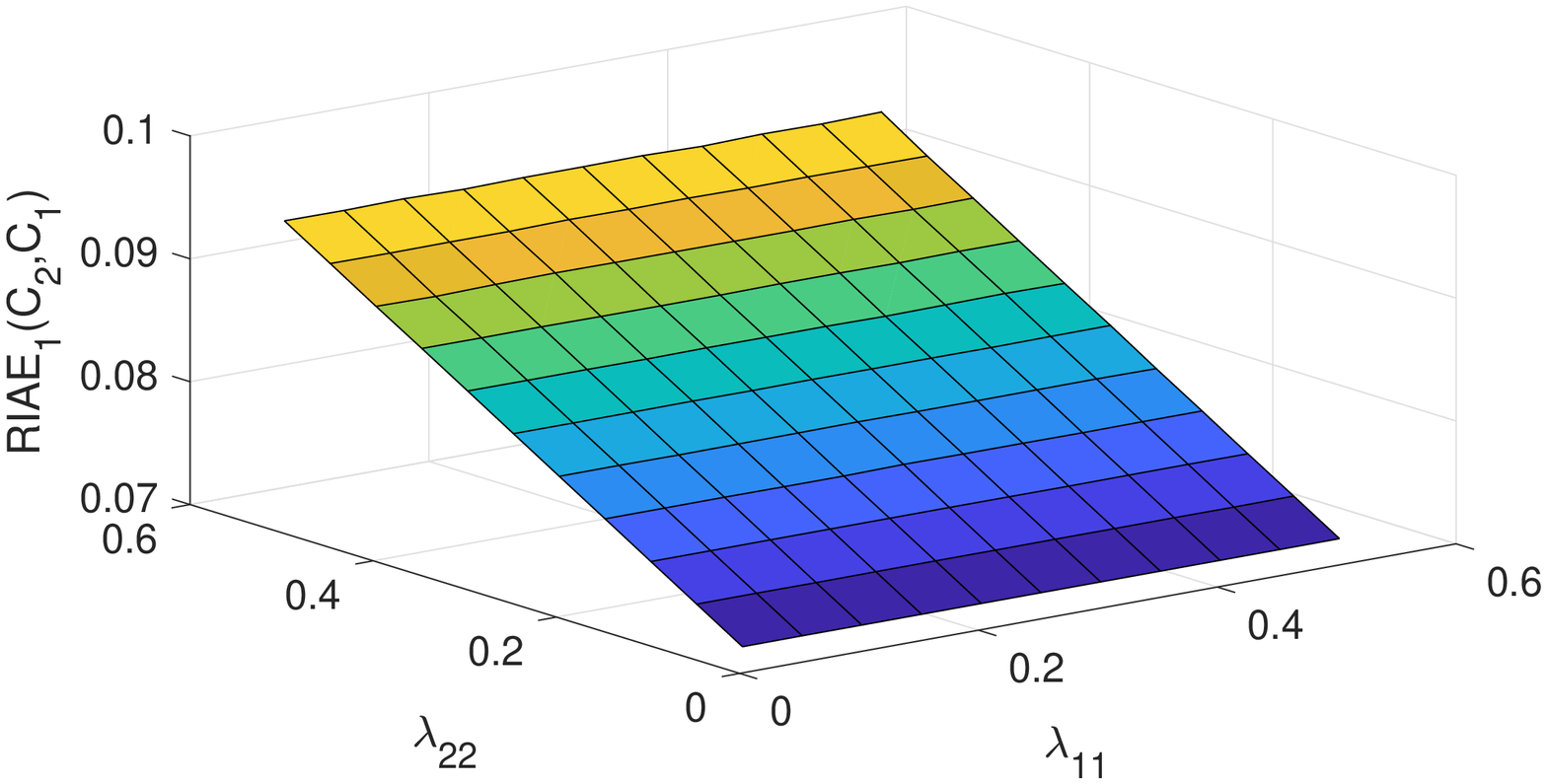}}
	\end{center}
	\caption[caption]{$RIAE$ indices surface for a) PID-IMC $G_{11}$ and b) PID-IMC $G_{22}$}
	\label{fig9}
\end{figure}
The $RITAE$ index for the controller PID-IMC $G_{11}$ is presented in Fig. 10a, which correspond to  the first trajectory change in $T_{e,sec,out}$. As can be observed, the linear shape of the surface coincide with the optimal point $\lambda_{11}=0.01$ and $\lambda_{22}=0.11$. Likewise, Fig. 10b shows the $RITAE$ index for the controller PID-IMC $G_{22}$, showing that the optimal point of Fig. \ref{fig8} coincides with the lowest value of this surface. The same optimal point is presented when the trajectory changes in $T_{e,sec,out}$ as can be observed in Fig. 11a and Fig. 11b.
\begin{figure}
	\centering
		\subfloat[]{\includegraphics[width=0.25\textwidth,height=0.09\textheight]{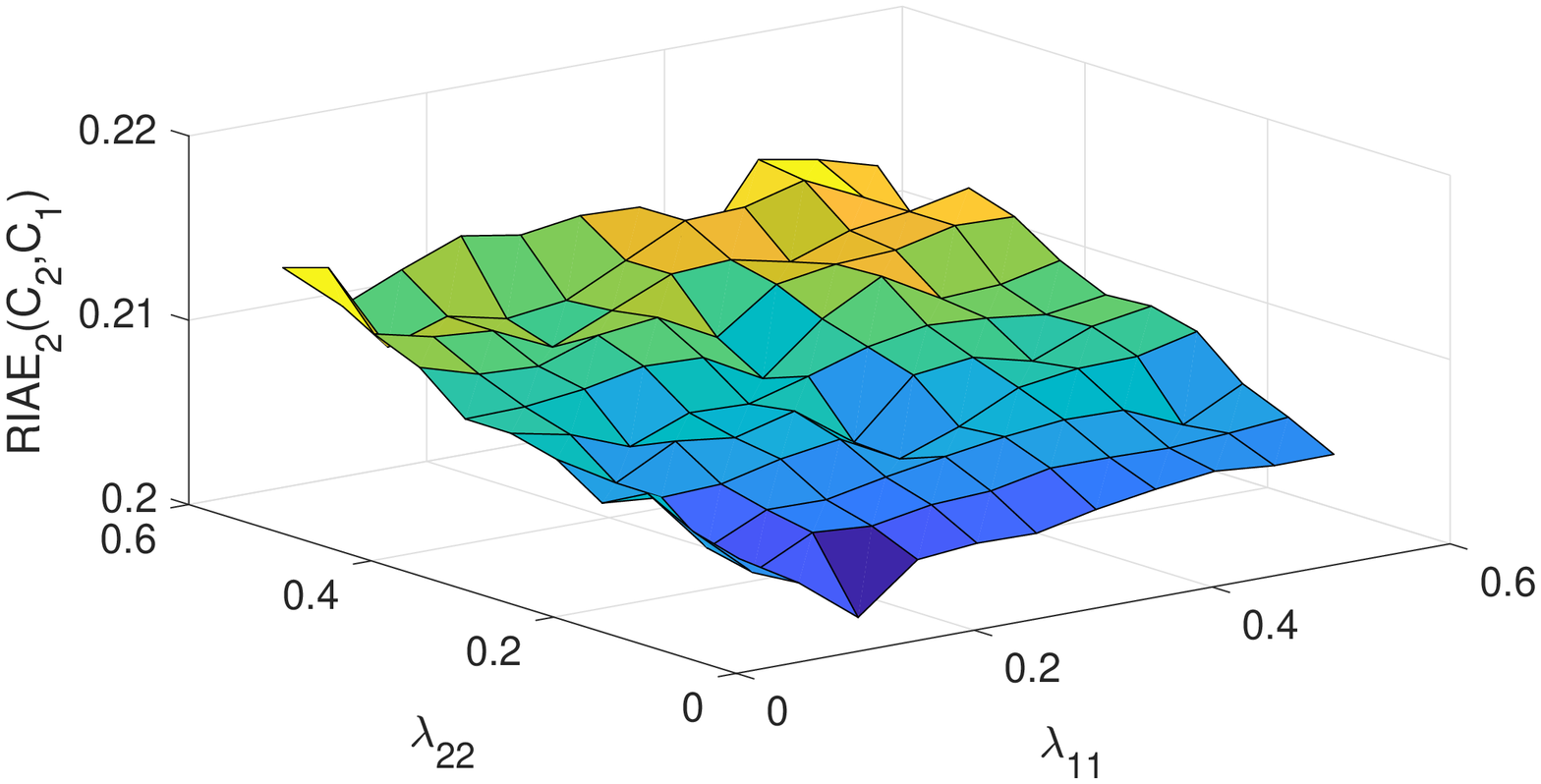}} 
		\subfloat[]{\includegraphics[width=0.25\textwidth,height=0.09\textheight]{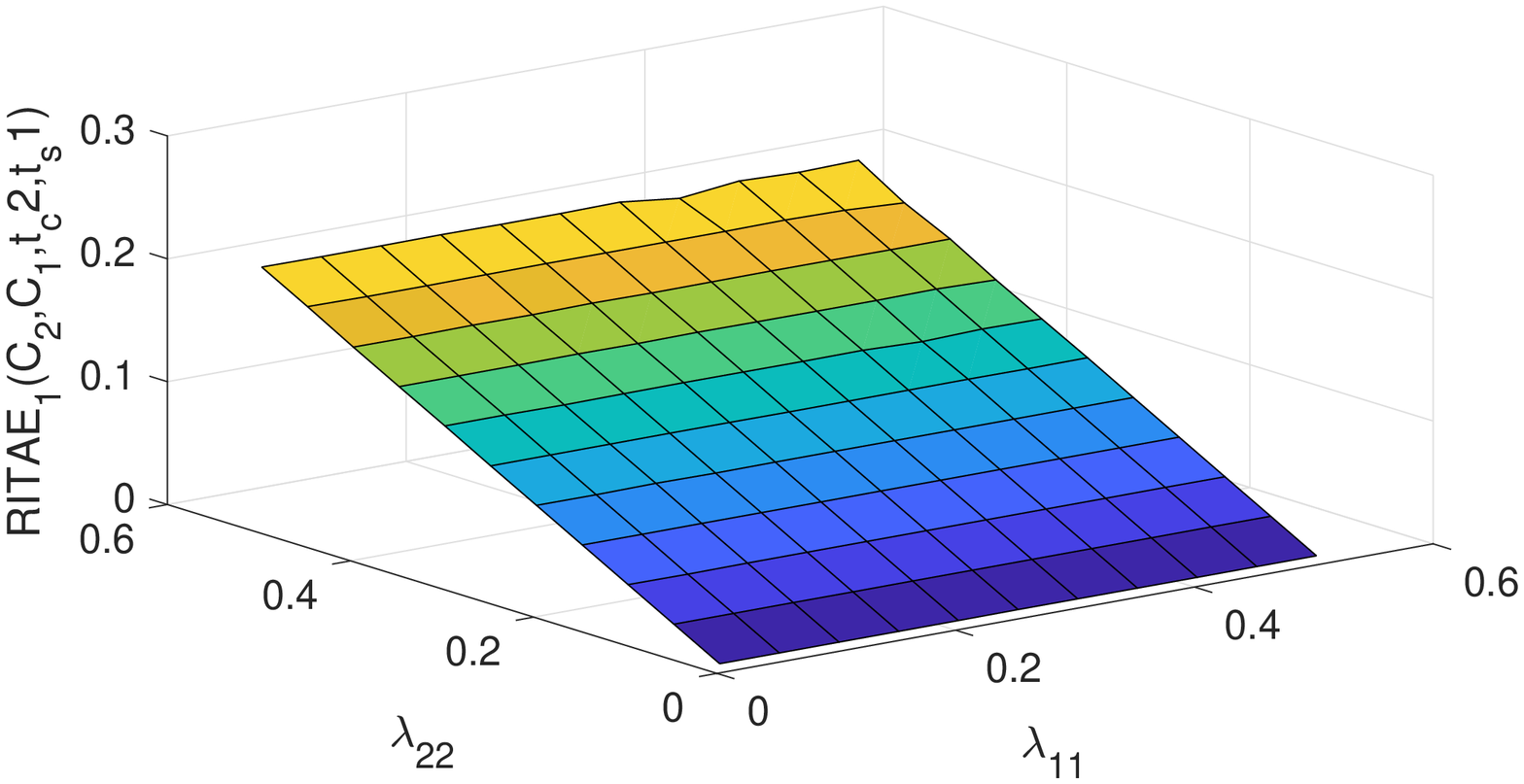}}
	\caption[caption]{$RITAE$ indices surface for a) PID-IMC $G_{11}$ and b) PID-IMC $G_{22}$}
	\label{fig10}
\end{figure}
\begin{figure}
	\centering
		\subfloat[]{\includegraphics[width=0.25\textwidth,height=0.09\textheight]{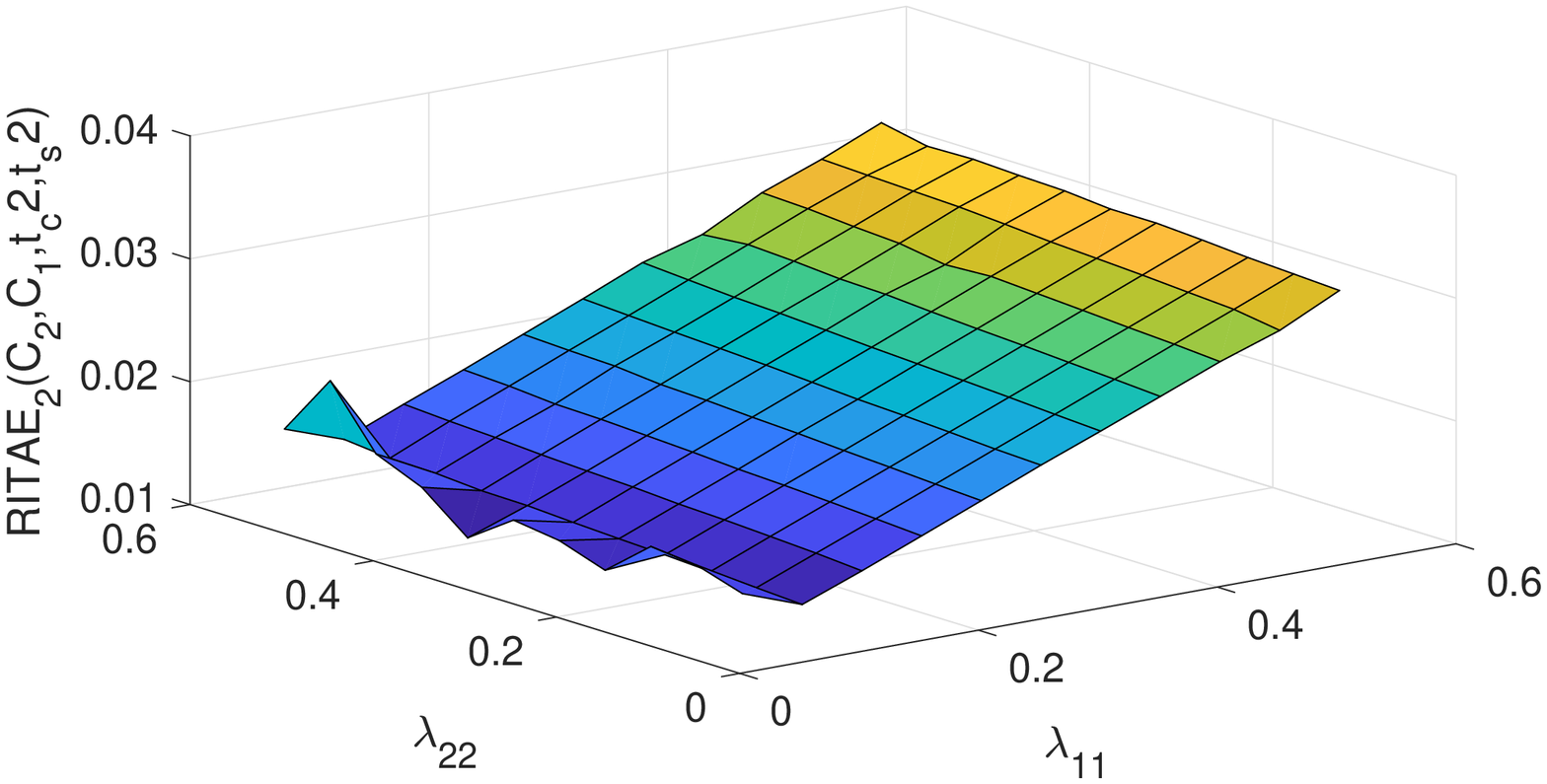}} 
		\subfloat[]{\includegraphics[width=0.25\textwidth,height=0.09\textheight]{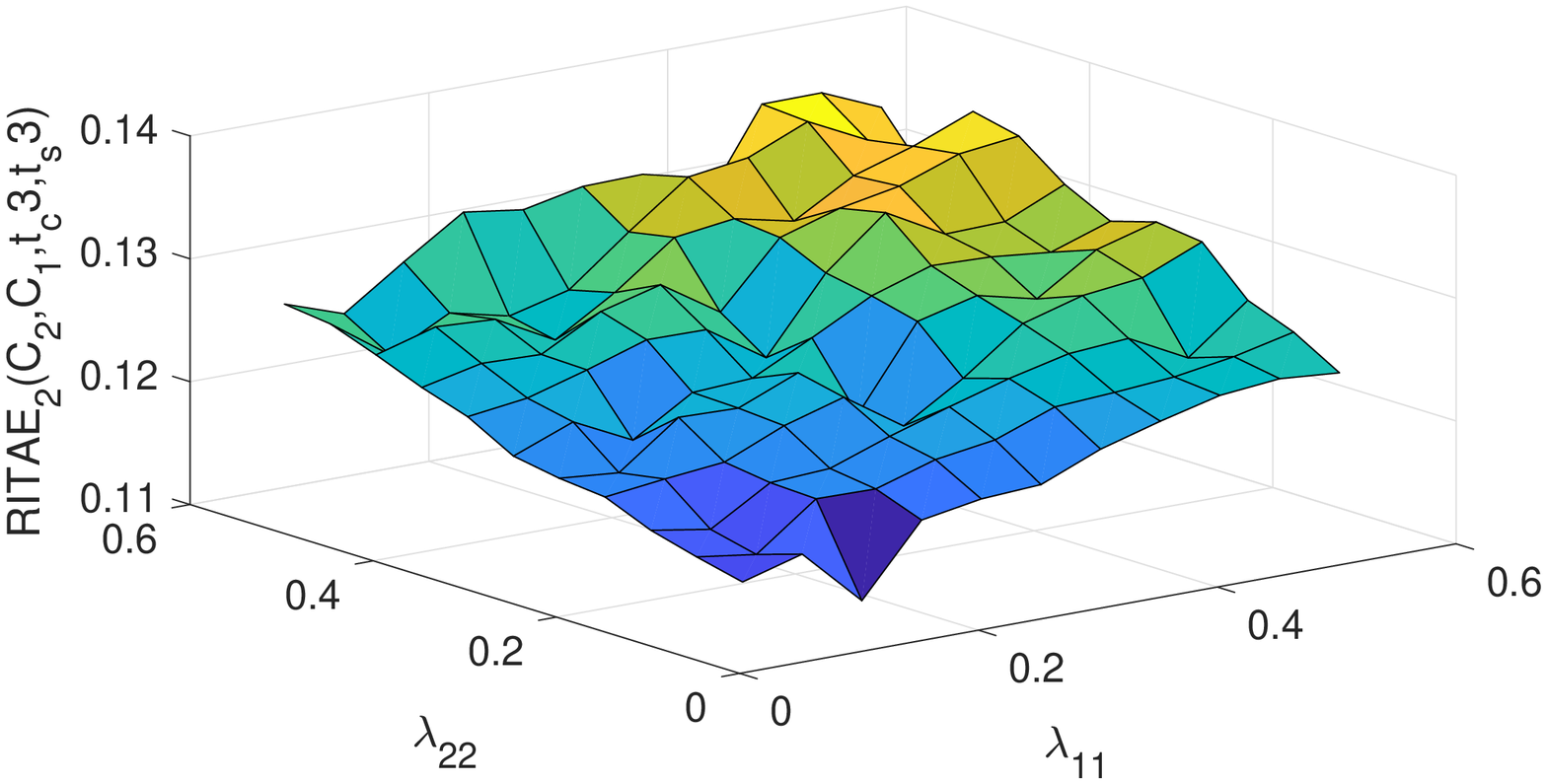}}
		\caption[caption]{$RITAE$ indices surface for PID-IMC  $G_{22}$ at the a) second and b) third trajectory changes}
	\label{fig11}
\end{figure}
The surface for the $RIAVU$ index of the controllers PID-IMC $G_{11}$ and PID-IMC $G_{22}$ is presented in Fig. 12a and Fig. 12b respectively. As can be observed, for the optimal point determined in Fig. \ref{fig8}, the RIAVU index for both controllers has a lower value, but this value is not the lowest value of the surface. However, smaller values of the surface does not ensure the best transient and trajectory performance of the system.
\begin{figure}
	\centering
		\subfloat[]{\includegraphics[width=0.25\textwidth,height=0.09\textheight]{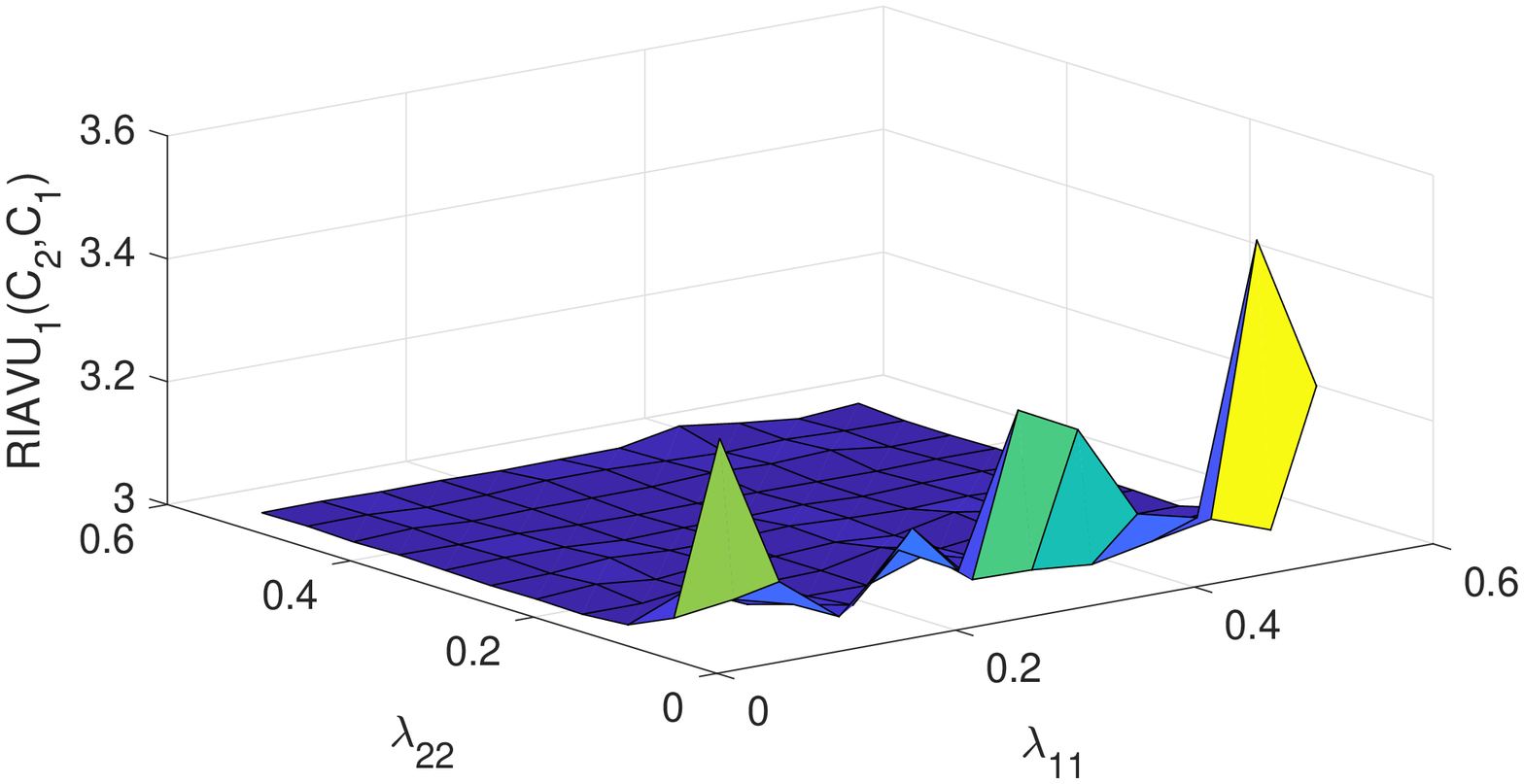}} 
		\subfloat[]{\includegraphics[width=0.25\textwidth,height=0.09\textheight]{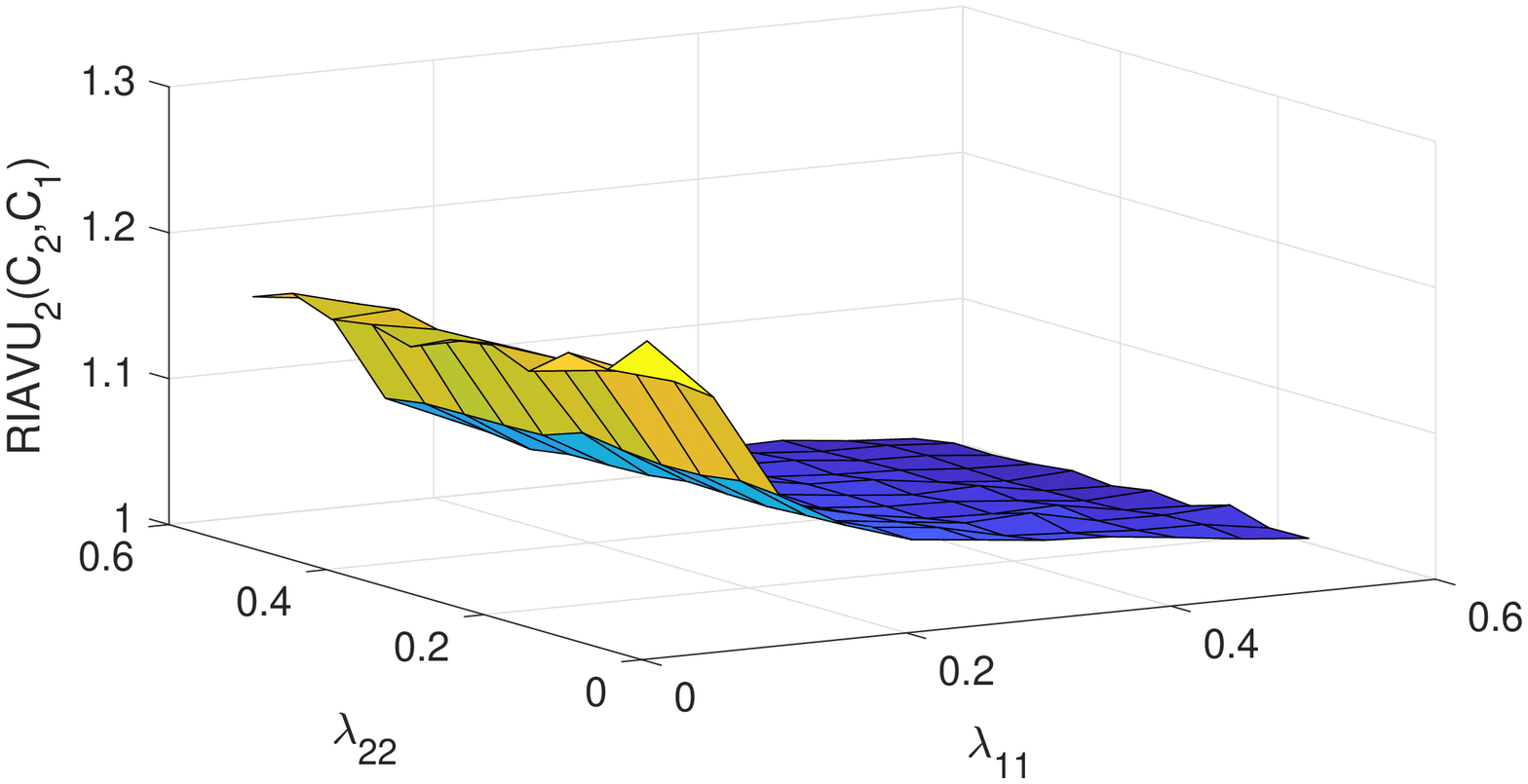}}
	\caption[caption]{$RIAVU$ index for a) PID-IMC $G_{11}$ and b) PID-IMC $G_{22}$}
	\label{fig12}
\end{figure}
\subsection{Comparison among PID18 benchmark results}
The PID18 benchmark challenge was taken by various researchers around the world for the PID 18 conference \cite{c12}-\cite{c29}. For this reason, Table VI presents a comparison among the different control strategies employed for the temperature control of the refrigeration system. This comparison includes the first author's last name and affiliation, the employed control technique as well as the resulting performance indices (12), sorted from the lowest to the highest $J$ index. As can be observed, the best control strategy is the RIOTS MPC controller because it presents the lowest $J$ index value. In the case of the original PID18 decentralized PID controller, it is located almost at the end of the table with a higher $J$ index value. On the other hand, the most accurated methods employed for the control of the refrigeration system consider not only optimization algorithms but also the multivariable features of the system. Besides, the PID-IMC controller presented in this paper system is the second control strategy with the best performance for the temperature control of the refrigeration system with a $J$ index closer to the RIOTS MPC controller. Notice that the PID structure of the PID-IMC controller allows an easier practical implementation of this strategy into an industrial environment. For this reason, the PID-IMC controller can be considered as a good cost-benefit alternative for the control of refrigeration systems.
\begin{table*}
	\setlength\abovecaptionskip{0\baselineskip}
	\setlength{\belowcaptionskip}{-5pt}
	\caption{Comparison between the PID18 benchmark challenge control strategies}
	\centering
		\small\addtolength{\tabcolsep}{-5pt}
		\begin{tabular}{|c|c|c|c|c|c|c|c|c|c|c|c|c|c|c|c|c|c|}		
			\hline
			\multirow{2}{*}{Pos}&Control&First &\multirow{2}{*}{University}&\multirow{2}{*}{Country}&\multirow{2}{*}{RIAE1}&\multirow{2}{*}{RIAE2}&\multirow{2}{*}{RITAE1}&\multirow{2}{*}{RITAE2}&\multirow{2}{*}{RITAE2}&\multirow{2}{*}{RITAE2}&\multirow{2}{*}{RIAVU1}&\multirow{2}{*}{RIAVU2}&\multirow{2}{*}{J}\\
			& technique& Author& & & & & & & & & & & 	\\
			\hline
			\multirow{2}{*}{\centering 1}&\multirow{1}{*}{RIOTS }&\multirow{2}{*}{\centering Dehghan,S}&\multirow{2}{*}{\centering UC Merced}&\multirow{2}{*}{\centering US}&\multirow{2}{*}{\centering 0.2134}&\multirow{2}{*}{\centering 0.1047}&\multirow{2}{*}{\centering 0.1943}&\multirow{2}{*}{\centering 0.008}&\multirow{2}{*}{\centering 0.012}&\multirow{2}{*}{\centering 0.0241}&\multirow{2}{*}{\centering 1.1481}&\multirow{2}{*}{\centering 1.0938}&\multirow{2}{*}{\centering 0.2055}\\
			&MPC&& & & & & & & & & & &\\
			\hline
			\multirow{2}{*}{\centering 2}&\multirow{1}{*}{PID-IMC}&\multirow{2}{*}{\centering Viola,J}&\multirow{2}{*}{\centering UC Merced}&\multirow{2}{*}{\centering US}&\multirow{2}{*}{\centering 0.071}&\multirow{2}{*}{\centering 0.20}&\multirow{2}{*}{\centering 0.003}&\multirow{2}{*}{\centering 0.016}&\multirow{2}{*}{\centering 0.1161}&\multirow{2}{*}{\centering 0.003}&\multirow{2}{*}{\centering 3.12}&\multirow{2}{*}{\centering 1.1}&\multirow{2}{*}{\centering 0.2065}\\
			&controller & & & & & & & & & & & &\\
			\hline
			\multirow{2}{*}{\centering 3}&\multirow{1}{*}{MIMO}&\multirow{2}{*}{\centering Tari,M}&\multirow{2}{*}{\centering Bordeaux U}&\multirow{2}{*}{\centering France}&\multirow{2}{*}{\centering 0.2819}&\multirow{2}{*}{\centering 0.2918}&\multirow{2}{*}{\centering 0.21776}&\multirow{2}{*}{\centering 0.09181}&\multirow{2}{*}{\centering 0.13766}&\multirow{2}{*}{\centering 0.065}&\multirow{2}{*}{\centering 1.0567}&\multirow{2}{*}{\centering 1.11143}&\multirow{2}{*}{\centering 0.2751}\\
			&robust PID & & & & & & & & & & & &\\
			\hline	
			\multirow{2}{*}{\centering 4}&\multirow{1}{*}{Data-driven}&\multirow{2}{*}{\centering Huff,D}&\multirow{2}{*}{\centering UFRS}&\multirow{2}{*}{\centering Brazil}&\multirow{2}{*}{\centering 0.1406}&\multirow{2}{*}{\centering 0.251}&\multirow{2}{*}{\centering 0.5444}&\multirow{2}{*}{\centering 0.0708}&\multirow{2}{*}{\centering 0.1901}&\multirow{2}{*}{\centering 0.03556}&\multirow{2}{*}{\centering 1.13}&\multirow{2}{*}{\centering 1.243}&\multirow{2}{*}{\centering 0.3225}\\
			&error& & & & & & & & & & & &\\
			\hline
			\multirow{2}{*}{\centering 5}&\multirow{1}{*}{Inverted}&\multirow{2}{*}{\centering Garrido,J}&\multirow{2}{*}{\centering Seville U}&\multirow{2}{*}{\centering Spain}&\multirow{2}{*}{\centering 0.158}&\multirow{2}{*}{\centering 0.258}&\multirow{2}{*}{\centering 0.536}&\multirow{2}{*}{\centering 0.095}&\multirow{2}{*}{\centering 0.186}&\multirow{2}{*}{\centering 0.113}&\multirow{2}{*}{\centering 1.338}&\multirow{2}{*}{\centering 1.431}&\multirow{2}{*}{\centering 0.353}\\
			&decoupling&& & & & & & & & & & &\\
			\hline
			\multirow{2}{*}{\centering 6}&\multirow{1}{*}{Feedforward}&\multirow{2}{*}{\centering Zhao,Y}&\multirow{2}{*}{\centering UC Merced}&\multirow{2}{*}{\centering US}&\multirow{2}{*}{\centering 0.4403}&\multirow{2}{*}{\centering 0.3297}&\multirow{2}{*}{\centering 0.3577}&\multirow{2}{*}{\centering 0.0812}&\multirow{2}{*}{\centering 0.111}&\multirow{2}{*}{\centering 0.1001}&\multirow{2}{*}{\centering 1.0738}&\multirow{2}{*}{\centering 1.5796}&\multirow{2}{*}{\centering 0.3744}\\
			&control&& & & & & & & & & & &\\
			\hline
			\multirow{2}{*}{\centering 7}&\multirow{1}{*}{Multiobjective}&\multirow{1}{*}{Reynoso-}&\multirow{2}{*}{\centering UPC}&\multirow{2}{*}{\centering Spain}&	\multirow{2}{*}{\centering 0.2892}&\multirow{2}{*}{\centering 0.3569}&\multirow{2}{*}{\centering 0.6148}&\multirow{2}{*}{\centering 0.1705}&\multirow{2}{*}{\centering 0.2291}&\multirow{2}{*}{\centering 0.0967}&\multirow{2}{*}{\centering 1.147}&\multirow{2}{*}{\centering 1.1531}&\multirow{2}{*}{\centering 0.4028}\\
			& Optimization& Mesa,G& & & & & & & & & & &\\
			\hline
			\multirow{2}{*}{\centering 8}&\multirow{1}{*}{MIMO virtual}&\multirow{2}{*}{\centering Bordignon,V}&\multirow{2}{*}{\centering UFRS}&\multirow{2}{*}{\centering Brazil}&\multirow{2}{*}{\centering 0.1569}&\multirow{2}{*}{\centering 0.3338}&\multirow{2}{*}{\centering 0.7193}&\multirow{2}{*}{\centering 0.3466}&\multirow{2}{*}{\centering 0.3704}&\multirow{2}{*}{\centering 0.0867}&\multirow{2}{*}{\centering 1.0499}&\multirow{2}{*}{\centering 0.9301}&\multirow{2}{*}{\centering 0.4134}\\
			&feedback& & & & & & & & & & & &\\
			\hline
			\multirow{2}{*}{\centering 9}&\multirow{1}{*}{Decentralized}&\multirow{2}{*}{\centering Zhang,B}&\multirow{2}{*}{\centering NCEPU}&\multirow{2}{*}{\centering China}&\multirow{2}{*}{\centering 0.3951}&\multirow{2}{*}{\centering 0.5881}&\multirow{2}{*}{\centering 0.241}&\multirow{2}{*}{\centering 0.4858}&\multirow{2}{*}{\centering 0.5805}&\multirow{2}{*}{\centering 0.122}&\multirow{2}{*}{\centering 0.9751}&\multirow{2}{*}{\centering 0.7346}&\multirow{2}{*}{\centering 0.4182}\\
			&ADRC& & & & & & & & & & & &\\
			\hline
			\multirow{2}{*}{\centering 10}&\multirow{1}{*}{Robust PID}&\multirow{2}{*}{\centering Zhao,S}&\multirow{2}{*}{\centering NCEPU}&\multirow{2}{*}{\centering China}&\multirow{2}{*}{\centering 0.3951}&\multirow{2}{*}{\centering 0.5881}&\multirow{2}{*}{\centering 0.241}&\multirow{2}{*}{\centering 0.4858}&\multirow{2}{*}{\centering 0.5805}&\multirow{2}{*}{\centering 0.122}&\multirow{2}{*}{\centering 0.9751}&\multirow{2}{*}{\centering 0.7346}&\multirow{2}{*}{\centering 0.4182}\\
			&Autotuning& & & & & & & & & & & &\\
			\hline
			\multirow{2}{*}{\centering 11}&\multirow{1}{*}{Model free}&\multirow{2}{*}{\centering Yu,X}&\multirow{2}{*}{\centering BJU}&\multirow{2}{*}{\centering China}&	\multirow{2}{*}{\centering 0.2289}&\multirow{2}{*}{\centering 0.3516}&\multirow{2}{*}{\centering 1.0495}&\multirow{2}{*}{\centering 0.3267}&\multirow{2}{*}{\centering 0.6223}&\multirow{2}{*}{\centering 0.1593}&\multirow{2}{*}{\centering 1.0029}&\multirow{2}{*}{\centering 0.903}&\multirow{2}{*}{\centering 0.5393}\\
			& adaptive control& & & & & & & & & & & &\\
			\hline
			\multirow{2}{*}{\centering 12}&\multirow{1}{*}{Conditional }&\multirow{2}{*}{\centering Yuan,J}&\multirow{2}{*}{\centering UC Merced}&\multirow{2}{*}{\centering US}&	\multirow{2}{*}{\centering 0.406}&\multirow{2}{*}{\centering 0.4043}&\multirow{2}{*}{\centering 0.3682}&\multirow{2}{*}{\centering 0.5573}&\multirow{2}{*}{\centering 0.3682}&\multirow{2}{*}{\centering 0.2244}&\multirow{2}{*}{\centering 1.7494}&\multirow{2}{*}{\centering 1.7494}&\multirow{2}{*}{\centering 0.5662}\\
			&Integrator& & & & & & & & & & & &\\
			\hline
	
			\multirow{2}{*}{\centering 13}&\multirow{1}{*}{IMC}&\multirow{2}{*}{\centering Cajo,R}&\multirow{2}{*}{\centering Ghent U}&\multirow{2}{*}{\centering Belgium}&	\multirow{2}{*}{\centering 0.8716 }&\multirow{2}{*}{\centering 0.8399}&\multirow{2}{*}{\centering 0.2954}&\multirow{2}{*}{\centering 0.6283}&\multirow{2}{*}{\centering 0.7651}&\multirow{2}{*}{\centering 0.3961}&\multirow{2}{*}{\centering 0.9024}&\multirow{2}{*}{\centering 0.7064}&\multirow{2}{*}{\centering 0.6332}\\
			&control& & & & & & & & & & & &\\

			\hline

			\multirow{2}{*}{\centering 14}&\multirow{1}{*}{Stochastic}&\multirow{2}{*}{\centering Ates,A}&\multirow{2}{*}{\centering UC Merced}&\multirow{2}{*}{\centering US}&	\multirow{2}{*}{\centering 0.8728 }&\multirow{2}{*}{\centering 0.8170}&\multirow{2}{*}{\centering 0.6820}&\multirow{2}{*}{\centering 0.3044}&\multirow{2}{*}{\centering 0.5517}&\multirow{2}{*}{\centering 0.1572}&\multirow{2}{*}{\centering 0.9755}&\multirow{2}{*}{\centering 0.9646}&\multirow{2}{*}{\centering 0.6532}\\
			&optimization& & & & & & & & & & & &\\
			\hline
			\multirow{2}{*}{\centering 15}&\multirow{1}{*}{Robust}&\multirow{2}{*}{\centering Rodriguez,D}&\multirow{2}{*}{\centering Seville U}&\multirow{2}{*}{\centering Spain}&\multirow{2}{*}{\centering 0.6046}&\multirow{2}{*}{\centering 0.6563}&\multirow{2}{*}{\centering 0.8932}&\multirow{2}{*}{\centering 0.5561}&\multirow{2}{*}{\centering 0.5043}&\multirow{2}{*}{\centering 0.4297}&\multirow{2}{*}{\centering 1.0012}&\multirow{2}{*}{\centering 0.7661}&\multirow{2}{*}{\centering 0.66}\\
			&Decoupling& & & & & & & & & & & &\\
			\hline
			\multirow{2}{*}{\centering 16}&\multirow{1}{*}{PID18 }&\multirow{2}{*}{\centering Bejarano,G}&\multirow{2}{*}{\centering Seville U}&\multirow{2}{*}{\centering Spain}&\multirow{2}{*}{\centering 0.3511}&\multirow{2}{*}{\centering 0.4458}&\multirow{2}{*}{\centering 1.6104}&\multirow{2}{*}{\centering 0.183}&\multirow{2}{*}{\centering 0.3196}&\multirow{2}{*}{\centering 0.128}&\multirow{2}{*}{\centering 1.1283}&\multirow{2}{*}{\centering 1.3739}&\multirow{2}{*}{\centering 0.68209}\\
			&Benchmark& & & & & & & & & & & &\\
			\hline
			\multirow{2}{*}{\centering 17}&\multirow{1}{*}{Robust FO }&\multirow{2}{*}{\centering Muresan,C}&\multirow{2}{*}{\centering Ghent U}&\multirow{2}{*}{\centering Belgium}&\multirow{2}{*}{\centering 0.3609}&\multirow{2}{*}{\centering 0.5935}&\multirow{2}{*}{\centering 0.027}&\multirow{2}{*}{\centering 2.8161}&\multirow{2}{*}{\centering 1.6339}&\multirow{2}{*}{\centering 0.4625}&\multirow{2}{*}{\centering 0.927}&\multirow{2}{*}{\centering 0.6868}&\multirow{2}{*}{\centering 0.7837}\\
			&controller& & & & & & & & & & & &\\
			\hline
			\multirow{2}{*}{\centering 18}&\multirow{1}{*}{nonlinear}&\multirow{2}{*}{\centering Lei,Z}&\multirow{2}{*}{\centering CEST}&\multirow{2}{*}{\centering China}&\multirow{2}{*}{\centering 0.98}&\multirow{2}{*}{\centering 1.18}&\multirow{2}{*}{\centering 0.81}&\multirow{2}{*}{\centering 3.25}&\multirow{2}{*}{\centering 1.01}&\multirow{2}{*}{\centering 1.43}&\multirow{2}{*}{\centering 0.99}&\multirow{2}{*}{\centering 0.62}&\multirow{2}{*}{\centering 1.26}\\
			&PID controller& & & & & & & & & & & &\\
			\hline
			\multirow{2}{*}{\centering 19}&\multirow{1}{*}{Evolutionary}&\multirow{2}{*}{\centering Amador,G}&\multirow{2}{*}{\centering UNAL}&\multirow{2}{*}{\centering Colombia}&\multirow{2}{*}{\centering 0.752}&\multirow{2}{*}{\centering 0.6558}&\multirow{2}{*}{\centering 5.4358}&\multirow{2}{*}{\centering 0.4548}&\multirow{2}{*}{\centering 0.497}&\multirow{2}{*}{\centering 1.43}&\multirow{2}{*}{\centering 1.0658}&\multirow{2}{*}{\centering 1.1159}&\multirow{2}{*}{\centering 1.7472}\\
			&PID& & & & & & & & & & & &\\
			\hline
	\end{tabular}
\end{table*}
\section{Conclusions}
This paper presented the temperature control of a  $2\times2$ multivariable refrigeration system employing the internal model control technique. The control system employed two PID controllers tuned by the internal model control technique. The performance of the PID-IMC controllers was contrasted with a decentralized PID controller proposed in the benchmark PID18 employing quantitative performance indices. Both controllers were tested for tracking tasks in the presence of external disturbances. The obtained results showed that the PID-IMC controllers have better tracking performance than the decentralized PID controllers due to the PID-IMC controllers have a lower performance index $J$. Besides, the obtained results of the PID-IMC controller was contrasted with all the results of the PID18 benchmark challenge, showing that the proposed controller has a better response than most of the implemented control strategies. We can conclude that the PID-IMC controller can be considered as a reliable alternative to the control of refrigeration systems.
\section*{ACKNOWLEDGMENT}
The authors are thankful to the members of PTUC/CPC (precision temperature uniformity control / cognitive process control) research group at UC Merced MESA Lab: Jie Yuan, Yang Zhao and Sina Dehghan for their contribution to this work.

\addtolength{\textheight}{-12cm}   
	

\end{document}